\documentclass[runninghead]{llncs}

\usepackage{array}
\newcolumntype{L}[1]{>{\raggedright\let\newline\\\arraybackslash\hspace{0pt}}m{#1}}
\usepackage{longtable}
\usepackage{scalerel,fp,caption,xcolor}
\usepackage{float}

\usepackage{natbib}
\setcitestyle{numbers,open={[},close={]}}

\usepackage{graphicx}

\begin{document}

\title{A Checklist for Explainable AI in the Insurance Domain}

\author{Olivier Koster\inst{1}\orcidID{0000-0003-1233-7156}
\and \\
Ruud Kosman\inst{2,3}\orcidID{0000-0002-3748-4198}
\and \\
Joost Visser\inst{1}\orcidID{0000-0003-0158-3095}}
\institute{%
LIACS, Leiden University, The Netherlands\\
\email{j.visser@liacs.leidenuniv.nl} 
\and
SIVI, The Netherlands
\and
InnoValor, The Netherlands}

\maketitle  

\begin{abstract}
Artificial intelligence (AI) is a powerful tool to accomplish a great many tasks. This exciting branch of technology is being adopted increasingly across varying sectors, including the insurance domain. With that power arise several complications. One of which is a lack of transparency and explainability of an algorithm for experts and non-experts alike. This brings into question both the usefulness as well as the accuracy of the algorithm, coupled with an added difficulty to assess potential biases within the data or the model. In this paper, we investigate the current usage of AI algorithms in the Dutch insurance industry and the adoption of explainable artificial intelligence (XAI) techniques. Armed with this knowledge we design a checklist for insurance companies that should help assure quality standards regarding XAI and a solid foundation for cooperation between organisations. This checklist extends an existing checklist that SIVI, the standardisation institute for digital cooperation and innovation in Dutch insurance.
\end{abstract}

\section{Introduction}

Artificial intelligence (AI) is one of the leading technologies paving the way for more efficient solutions and powerful automation. This exciting technology is being deployed increasingly across various industries.
For instance, AI is aiding healthcare in its search for accurate diagnostic procedures in order to detect cancer early, assisting radiology by discovering patterns and accelerating medicine development \citep{builtin}. Similarly, in the insurance industry, the use of AI is starting to gain traction, being used in assessing risks, handling claims and detecting fraud. Aside from all this added ability, AI too comes with its own downsides. Much like human cognition, technology has its flaws. The same goes for AI algorithms. 
While early AI systems were relatively easy to comprehend, we have seen a recent rise in opaque decision systems such as Deep Neural Networks. Although these types of algorithms increase accuracy, they come with a higher level of algorithmic complexity, often consisting of hundreds of layers and millions of parameters \citep{Castelvecchi}. In these instances, interpretability vastly decreases. This results in a black-box algorithm that is difficult to understand for experts and non-experts alike. This can create some unfavourable situations. In particular, when decisions made by an algorithm affect human lives. For example, when black-box algorithms make incorrect diagnoses, doctors may be subject to intense scrutiny for taking the wrong course of action, being unable to explain the proper reasoning behind a diagnosis. This phenomenon can have an even more severe impact on a larger scale. In 2020, The Dutch government deployed SyRi, an algorithmic fraud risk scoring system. It used a non-disclosed algorithmic risk model to profile citizens, allegedly targeting mostly low-income neighbourhoods and minority residents. Dutch court deemed SyRi illegal for lacking transparent data usage and violating privacy. Similarly, Apple revealed its AI driven credit risk system showed a striking bias, as it deemed men far more creditworthy than women. Consequently, individuals could be given different credit limits despite having the same accounts, cards or assets \citep{factAI}. In precision medicine, decisions cannot be based on mere binary prediction, creating a need for extensive explanations supporting a model's output. The same holds true for other domains such as autonomous vehicles, transportation, security, and finance \citep{MedicalXAI}.

Evidently, using AI to make impactful decisions can be a dangerous practice when the legitimacy of the model is not justified. From this perspective, the responsibility of the algorithms' creator does not only concern its accuracy but also  its interpretability and transparency. This reasoning has spawned a new field of research named Explainable Artificial Intelligence (XAI). Research on this topic has spiked in recent years, reflecting a growing need for XAI. Even so, to our knowledge, no prior research captures the state of AI adoption, and more specifically of XAI techniques, in the (Dutch) insurance industry. 

In this paper, we examine the current adoption and future prospects of AI and XAI within the Dutch insurance industry. This is done through literature research coupled with conducting exploratory interviews of industry experts. All of this will lay the groundwork for the design of a checklist for insurance companies that should help assure quality standards regarding XAI and a solid foundation for cooperation between organisations. The checklist is evaluated and tested by conducting confirmatory interviews, creating a feedback loop for further refinement. The aforementioned checklist extends an existing checklist regarding AI that SIVI, the standardisation institute for digital cooperation and innovation in Dutch insurance.

\section{Method}\label{method}

Our research methodology is based on the design science research paradigm for information systems \citep{DesignResearch}. Contrary to behavioural science, design science is focused on designing an artefact. In our case, this artefact is a checklist for explainable and transparent AI applications. Similar to the methodology discussed in \citep{DesignResearch}, research is done in three iterative cycles.

We conduct two types of interviews during our research. Firstly, we perform semi-structured exploratory interviews to assess the current state of the art with respect to AI techniques used within the insurance industry. This is helpful because, while current literature tells us a lot about the possibilities and practices of AI, it gives little insight into the actual adoption of AI techniques. Additionally, practical context is usually missing in most available literature. We interview four industry experts at companies that operate within the insurance industry, ranging from software suppliers to insurance companies. The findings can be found in Section \ref{exploration}. Secondly, we conduct confirmatory interviews with similar companies to evaluate our findings and checklist design. This will give us a strong indication of both the correctness, robustness as well as practicality of our design. Similarly to the exploratory interviews, evaluation is done by conducting four confirmatory interviews with industry experts from insurance companies, and their software suppliers. The evaluation findings can be found in Section \ref{validation}. For privacy reasons we will not disclose the actual company names, nor will we disclose employee names and information that could be used to identify organisations and personnel. Instead, we will call the seven companies `company A' through `company G', respectively. Note, that with company B we have conducted both an exploratory interview as well as a confirmatory interview.

\section{Literature Background}\label{related work}
\subsubsection*{Taxonomy of XAI}\label{Taxonomy}

According to \citet{XAISystemicReview}, there is little agreement among scholars on what explanations are and what properties they might have, as well as the correct terminology that should be used. Fortunately, \citet{TaxonomyXAI} have comprised a clear report on the most common terms used in XAI research and their respective meaning. Notably, they propose three distinct \textit{levels of transparency}. \citet{TaxonomyXAI} explain how these levels of transparency apply to different AI algorithms, enabling them to categorise the algorithms as either of type \textit{transparent} or type \textit{opaque} (or \textit{non-transparent}). Of course, a system itself is never opaque but rather opaque with respect to some particular agent. \citet{ToWhom?} break down all different agent groups within a system in fine detail. Depending on what a particular agent is tasked with doing, they are likely to require a different kind of knowledge to do it and, thus, seek a different kind of explanation. \citet{TaxonomyXAI} have comprised a list of the research goals that particular agents can achieve through model explainability. To get a better grip on the concept of explaining AI models, we must acknowledge that explanations come in many different proverbial shapes. \citet{XAISystemicReview} describe attributes and characteristics of explanations as well as define several types of explanations. In recent years, several XAI methods have been developed to increase model transparency and explainability. \citet{XAISystemicReview} have also compiled a list of all current \textit{ante-hoc} and \textit{post-hoc} XAI methods.

\subsubsection*{Responsible AI}\label{responsibleAI}

AI bias is a phenomenon that occurs when an algorithm produces results that are systemically prejudiced due to either erroneous (or correct but undesired) assumptions in the machine learning process. This can happen with learning algorithms when they are trained on their dataset. In this case, we call the phenomenon \textit{algorithmic bias}. Even so, other forms of bias exist. It can also result from human errors (e.g. faulty collection or representation of input data). \citet{Bias} break down all possible cause for bias. Sometimes biases can be functionally correct, but immoral to base results upon. These undesired biases stem from a black-box model's tendency to unintentionally create unfair decisions by including sensitive factors such as the individual's race or gender. This phenomenon gives rise to certain discriminatory issues, either explicitly (considering sensitive attributes) or implicitly (considering factors that correlate with sensitive attributes). \citet{TaxonomyXAI} gives several degrees of fairness that should be considered to design responsible AI solutions. 

Several guidelines for trustworthy AI have been proposed \cite{DNB,ainow}. Similar to the checklist we propose, these frameworks aim to give general guidance for responsible use of AI. However, these frameworks, while being very high level in nature, are not actionable for working developers. \citet{serban2021practices} provide 14 engineering practices for trustworthy ML applications, of which several are related to explainability. These practices are not formulated in terms of checks.

\section{Exploration}\label{exploration}

As mentioned, we carried out four exploratory interviews with industry experts from financial institutes, insurance companies, and their software suppliers. A summary of the interviewees is shown in table \ref{summaryexploratoryinterviewcompanies}. The priority of these interviews lies in achieving three primary goals. Firstly, we seek a more in-depth understanding of the insurance industry and its related processes, stakeholders, demands and concerns. Secondly, we want to know which AI techniques said companies deployed and which they plan to deploy in the future. Thirdly, we want to know how the industry values transparency and explainability.

\setlength{\arrayrulewidth}{0.2mm}
\setlength{\tabcolsep}{5pt}
\renewcommand{\arraystretch}{1.4}

\begin{table}
\caption{Summary of exploratory interview company information \\ $^{*}$Also takes part in confirmatory interview}
\label{summaryexploratoryinterviewcompanies}
\begin{tabular}[H]{@{}L{1.8cm}|L{2.2cm}L{2.2cm}L{2.2cm}L{2.3cm}@{}}
\hline
    & company A & company B$^{*}$ & company C & company D \\
\hline
core \hspace{5mm} business & insurance software supplier & insurance \hspace{5mm}intermediary \hspace{5mm} (software) & financial \hspace{5mm} services &  financial \hspace{5mm} services \\
 \# employees & 100+ & 100+ & 1.000+ & 10.000+ \\
 job-title interviewee(s) & product owner & ML engineer \& software engineer & manager client contact financial services & Product Owner \& Innovation \hspace{5mm} Manager \hspace{10mm}  Business  \hspace{5mm}  Automation 
 \\
 uses RBS for & various \hspace{5mm} processes & various \hspace{5mm} processes & various \hspace{5mm} processes & various \hspace{7mm} processes \\
 uses ML for & - & fraud detection & - & fraud detection \\
 uses DL for & - & - & - & - \\
 \hline
\end{tabular}
\end{table}
\subsubsection*{Findings}\label{Exploratoryinterviewfindings}

The insurance industry is still in a preliminary phase when it comes to the deployment of AI technologies. All of the four companies interviewed have deployed some form of Rule-Based System (RBS), but some are hesitant to adopt more complex AI techniques, like Machine Learning (ML) and Deep Learning (DL). This is because most companies are focused on improving these Rule-Based Systems and ironing out any inefficiencies. Additionally, the interviewees from company B expressed a growing concern that people lack trust in ML algorithms. It takes time and effort to convince people that ML algorithms work better than Rule-Based Systems, even though oftentimes they are statistically proven to do so. Thus, at the moment, all of them prefer Rule-Based Systems because they are more explainable, even when the results that are explained are sometimes less accurate. Two out of four companies interviewed have already deployed ML algorithms in their processes. Notably, both company B and D have deployed the technology for insurance claim handling, specifically to detect possible fraudulent activity. It is interesting to note, however, that both companies do so with different kinds of algorithms. Company B uses a decision tree classifier (supervised learning), whereas company D uses a k-means clustering algorithm for anomaly detection (unsupervised learning). Company B also uses ML algorithms for calculating car insurance premiums. They are also experimenting with ML algorithms to calculate a customer's risk coverage ratio, but this system has yet to be deployed. Importantly, neither use non-transparent algorithms (including DL) for these tasks. Company B and D are already experimenting with DL. However, none of the interviewed companies have deployed any DL systems thus far. There are four main reasons for this:
\begin{enumerate}
    \item \textbf{DL is less explainable}: Even though DL algorithms are usually more accurate than ML algorithms (and Rule-bases Systems), they are even less explainable. A balance between accuracy and explainability has to be found. Insurance companies mostly choose explainability in favour of a marginal increase in accuracy. They need increased explainability to understand and convey why the system gives a certain output, otherwise, the results are not actionable.
    \item \textbf{Understanding DL requires technical/mathematical expertise}: Because DL algorithms (and some ML algorithms for that matter) have such high complexity, they are less explainable. Thus, they require more expertise to be understood and used (this is the case for all system agent roles, but especially operators and executors). For most current employees this creates a knowledge gap that is hard to overcome. Furthermore, if they were to overcome this obstacle, their job description would change significantly. Claims handlers would turn into model experts.
    \item \textbf{DL is less transferable}: DL algorithms are sometimes less transferable than other AI techniques. The input data that insurance companies use at the moment is less suitable for these types of algorithms. Additionally, some of the data that could be used to extract the most out of DL algorithms are not present in the dataset or are off-limits due to privacy concerns.
    \item \textbf{Streamlining RBS has more value in the short term}: More quality gain can be found in streamlining current RBSs and ML processes, instead of looking for accuracy gain with DL algorithms. Gains can especially be made in the refinement of input data (e.g. feature selection), as this is where most resource and thought is going at the moment.
\end{enumerate}

\noindent
Future prospects for all companies range from the initial deployment of ML to the deployment of complex DL algorithms when the aforementioned issues start to be resolved. Most expect to start incorporating (more of) these complex AI technologies near the end of the 2025.

\section{Design}

\subsubsection*{Purpose}\label{Purpose}

The checklist should be a list of `checks' that, if answered properly, should test the explainability and transparency of AI model applications, as well as highlight potential weaknesses and areas for improvement. We define a \textit{check} as a component that features one or more questions, hence the collective is called a checklist. Every check comes with an elucidation to clear any confusion for the reader and to make sure the question is answered as intended. Checks either have open answers or multiple choice answers. The complete checklist can be found in the appendix. 
The checklist is designed with two main purposes in mind: Firstly, it should be used to confirm the quality and completeness of an AI application with regards to its explainability and transparency. In that way, the checklist can essentially be used as a guide to evaluate if all facets, that make a well designed explainable and transparent AI application, are accounted for. If, based on this checklist, one would conclude their application is not complete or lacks quality in certain key areas, it serves as an indication where further progress should be made. Secondly, the checklist, if properly filled in, could be shared with third parties (clients or companies) to show the quality and completeness of an AI applications with regards to explainability and transparency. This is especially helpful for collaboration between companies to give confidence that certain information or assets can be shared. Moreover, this could be interesting from a marketing standpoint, giving clients assurance that your application is well designed and responsible.

\subsubsection*{Constraints}\label{Constraints}

\renewcommand{\footnoterule}{
  \kern -3pt
  \hrule width \textwidth height 1pt
  \kern 2pt
}

To fulfil these purposes we formulate several constraints for the checklist design. The checklist is based on an existing checklist\footnote{The Checklist-KOAT can be found at \url{https://www.sivi.org/checklist-koat/}} named `Checklist-KOAT' or `Checklist Kwaliteit Onbemenste Advies- en Transactietoepassingen' by SIVI, the standardisation institute for digital cooperation and innovation in Dutch insurance. This existing checklist covers several topics with regards to computer applications for financial advice and financial transactions. We can deduce several helpful constraints that are implied in this pre-existing checklist. We will use these implied constraints as well as our design guidelines to set constraints to design our checklist. The following constraints are used:

\begin{enumerate}
  \item \textbf{Practical relevance}: We want our checklist to be applicable for practical use. That means that all covered topics should be relevant from a practical standpoint. Furthermore, the checklist cannot be overly long or be too technically in-depth, as this would disincline people from using it.
  \item \textbf{Non-expert terminology}: The checks and elucidation should refrain from using expert terminology as much as possible. If used in a practical environment by actual employees of financial companies, expert terminology may be unclear and would not induce a full understanding of the covered topic.
  \item \textbf{Broadness-precision balance}: Topics should be covered broadly enough to be appropriate for most, if not all, AI model applications. Yet, checks should be precise enough to get the most informative answer. A proper balance should be found between these two ends.
\end{enumerate}

\section{Validation}\label{validation}

As mentioned in Section \ref{method}, similarly to the exploratory interviews, evaluation is done by conducting four confirmatory interviews with industry experts from insurance companies, and their software suppliers. A summary of the interviewees is shown in table \ref{summaryconfirmatoryinterviewcompanies}. Other than to evaluate our design, these interviews essentially helped us confirm whether our original findings are correct and if they still hold within new contexts. The structure of the interview is as follows: For every check and its elucidation, we ask three things: 1. ``Is the phrasing and meaning clear?" 2. ``How relevant is the check and its encompassing topic (with regards to the purposes mentioned in Section \ref{Purpose})?" and 3. ``What would your answer be to the question for your specific application?" After all, topics are covered, we ask two general questions about the entire set of topics: 1. ``Do you deem the sequence/order of topics logical and favourable?" 2. ``Is the set of topics (and checks) complete or do you think a topic is missing?"

\setlength{\arrayrulewidth}{0.2mm}
\setlength{\tabcolsep}{5pt}
\renewcommand{\arraystretch}{1.4}

\begin{table}
\caption{Summary of confirmatory interview company information \\ $^{*}$Also takes part in exploratory interview}
\label{summaryconfirmatoryinterviewcompanies}
\begin{longtable}{@{}L{1.8cm}|L{2.2cm}L{2.2cm}L{2.2cm}L{2.2cm}@{}}
\hline
\raggedright & company E & company F & company G & company B$^{*}$ \\
\hline
core \hspace{5mm} business & software \hspace{5mm} supplier & insurance \& pensions & software supplier (insurance \& pensions) & insurance \hspace{5mm} intermediary (software)  \\
%\# 
\# employees & 10+ & 1.000+ & 10+ & 100+ \\
job-title interviewee(s) & CCO\textbackslash CMO & sr. IT Architect & user  interaction designer & ML engineer \& software engineer \\
%revenue$_{ (2020)}$ & \\
use RBS for & various \hspace{5mm} processes & various \hspace{5mm} processes & various \hspace{6mm} processes & various \hspace{5mm} processes  \\
use ML for & - & - & policy recommendation & fraud detection \\
use DL for  & -  & -  & -  & -  \\
\hline
\end{longtable}
\end{table}

Conforming to our used design science research methodology, evaluation is done during the design phase. Therefore, the design process has an iterative nature. Consequently, a new checklist draft is designed after each confirmatory interview. This way the design is improved in a step by step manner. Most initial constructive criticism, in the interview with company E, was aimed at phrasing and meaning (of checks and their elucidation) being unclear. This resulted in the inclusion of additional elucidations were needed, or rephrasing of said unclear pieces of information. This was the case throughout the design. In most cases, an illustrative example was also added in an attempt to clear up any remaining confusion. The next iteration was found to be much more clear and comprehensible, although slight improvements kept being made from version to version. Until, in the last interview, no confusion was remarked explicitly.

\section{Discussion}

In the final iteration of the design, all checks and elucidations seemed to be phrased clearly, to be fully understood by the interviewees, based on our assessment of their answers. Also, based on the results, all checks and topics present in the final iteration seemed to be relevant enough to be included in the design. Interviewees specifically expressed relevance for the topics spanning bias. Given this fact, more checks could be added towards this topic. Such questions could dive deeper into why they include and exclude certain biases in their model (thus, revealing which biases they would label as undesired biases). Eventually, we landed on a design that puts a heavy emphasis on questions formulated with open-ended answer in mind. This has two main advantages, whilst also running the risk of some potential drawbacks. The first advantage is that phrasing the questions in such a way, tends to squeeze as much interesting information out of a single check as possible, as long as the checklist user is motivated to explore the answer to the intended extent (interviewees have at least expressed the intention to do so). The second advantage is that this open-ended phrasing creates room for a certain broadness in the scope of a check's applicability. By restricting the answers too much, you run the risk of excluding some AI applications, rendering the check useless for their specific model.

At the end of the day, the checklist needs to be relevant for companies that only use RBSs, but must also be a guide to ML and DL technology, in order to aid with applications in the future. Based on the results of our confirmatory interviews, we feel our design fulfils this ambition.

\section{Conclusion}

\subsubsection*{Contributions}
Several contributions stand out when compared to other literature that we could find on the topics of XAI and AI in general.
Firstly, we propose a checklist that can be used to assess and help assure transparency and explainability for AI application in a practical environment. It can also be used to verify if enough thought has gone into the application and to share quality standards across parties. As such, our checklist is more actionable for the working developer than the frameworks mentioned in Section \ref{responsibleAI}.
Secondly, we give insight into AI and XAI adoption in the insurance industry. Few other papers talk about AI and XAI in the financial sector. Presumably, this comes down to the fact that most financial companies are only now starting to adopt AI algorithms effectively, as knowledge on the subject has only started to grow in recent years. After all, once a technology has been discovered, it takes some time for it to develop into a commercially viable product. Thirdly, we have, to some extent, validated existing theories and concepts about XAI in a practical environment.

%\subsection{Future work}
\subsubsection*{Future work}
As with any study, some things could be done to further improve the design research carried out. For example, while interviews were conducted to learn about the adoption and prospects of AI and XAI techniques in the Dutch insurance industry as a whole, more interviews would give a more complete view of the industry. Additionally, since industry experts employed at insurance companies and software suppliers were targeted for the interviews, end-customers, consumers and lawmakers were not consulted. Moreover, only a small number of companies in the industry are now starting to gain traction with ML and DL concepts. This means that knowledge of the technologies among industry experts is still relatively scarce. Considering, that in the future this knowledge will grow, more detailed analyses could be done on the topic.

Finally, as initially mentioned, the checklist is meant to extend an existing checklist, named `Checklist-KOAT', which is made by SIVI. Specifically, the design mentioned in this paper serves as a base for the eventual integration into the `Checklist-KOAT'. This integration will be done by SIVI itself. SIVI will keep improving the integrated design through field testing with associated member companies. We presume the design will remain relevant for the foreseeable future, although, as time progresses and new techniques become prevalent, eventual updates will inevitably be advisable.

While developed and validated in the context of the insurance domain, our checklist can likely be generalized to other domains.

\pagebreak

\appendix

\section*{Appendix: Checklist for AI in insurance applications}\label{Checklist}

\noindent\makebox[\textwidth]{%
\includegraphics[width=1.0\textwidth]{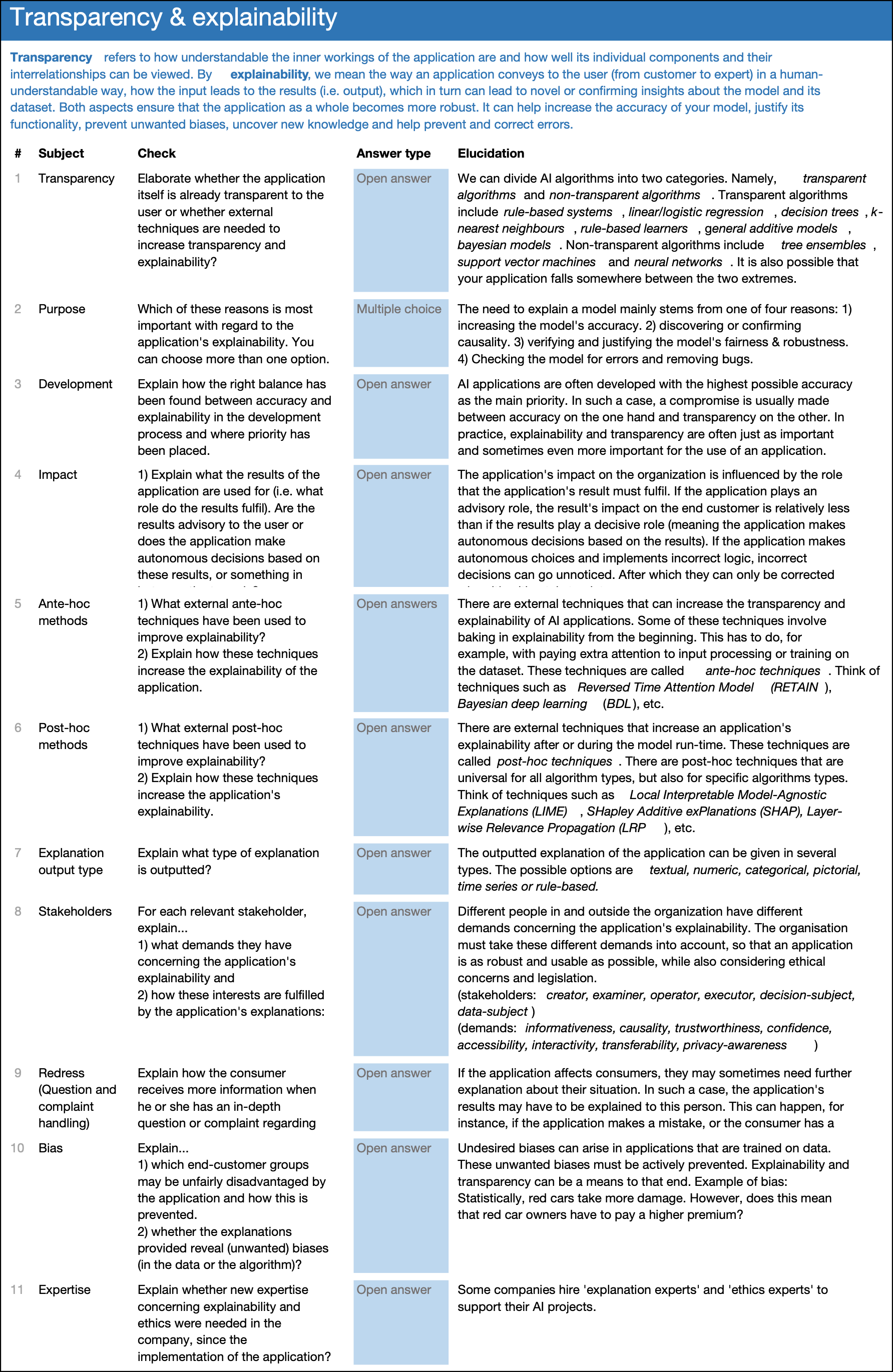}}

\bibliographystyle{splncs04nat}
\bibliography{paper}
\end{document}